\documentstyle[natbib_rich,psfig,epsf,pslatex]{mn}

\newcommand{\mgc}{\mbox{MCG$-$6-30-15 }}

\newcommand{\fluxplot}{\mbox{$\rm F_{h}$ vs $\rm F_{s}$ }}
\newcommand{\linmo}{\mbox{$F_{h}=kF_{s}+C$}}
\newcommand{\chisq}{$\chi^{2}$}

\title[X-ray spectral variability in AGN]{The nature of X-ray spectral
variability in Seyfert Galaxies}
\author[Taylor, Uttley \& M$^{\rm c}$Hardy]{Richard D.~Taylor\thanks{e-mail:
rdt@astro.soton.ac.uk}, Philip~Uttley and Ian~M.~M$^{\rm c}$Hardy
\\ Department of Physics and Astronomy, University of Southampton, Southampton, SO17 1BJ}
\date{Received ** *** 2003 / Accepted ** *** 200*}

\pagerange{\pageref{firstpage}--\pageref{lastpage}} \pubyear{200*}

\begin{document}
\maketitle

\label{firstpage}

\begin{abstract}
We use a model-independent technique to investigate the nature of
the 2-15~keV X-ray spectral variability in four Seyfert galaxies and
distinguish between spectral pivoting and the two-component model for
spectral variability.  Our analysis reveals conclusively that the softening of the X-ray
continuum with increasing flux in \mgc and NGC~3516 is a result of
summing two spectral components:
a soft varying component (SVC) with spectral shape independent of flux and a
constant hard component (HCC).  In contrast, the spectral variability
in NGC~4051 can be well described by simple pivoting of one
component, together with an additional hard constant
component.  The spectral variability model for NGC~5506
is ambiguous, due to the smaller range of fluxes sampled by the data.  We investigate
the shape of the hard spectral component in \mgc and find that it
appears similar to a pure reflection spectrum, but requires a large reflected fraction
($R>3$).  We briefly discuss physical interpretations of the different modes of spectral
variability.

\end{abstract}


\section{Introduction}

Rapid X-ray variability is common in Seyfert galaxies
\citep{grn93} and suggests that the X-ray emission originates close to
the central black hole.  In a number of Seyfert galaxies, the 2-20 keV
continuum spectrum steepens with increasing
X-ray flux (\citealt{iwa96}; \citealt{la00}; \citealt{la03};
 \citealt{lee00}; \citealt{chaing00}; \citealt{done00}).  This
softening of the X-ray emission is commonly attributed to a change in shape of
a single continuum component, possibly due to the Compton cooling of the X-ray
emitting corona by the increased seed photon flux \citep{haa97}.
  
Recent spectral variability studies of 
\mgc with \textit{RXTE} \citep{imhpap98} and subsequently with \textit{ASCA}
\citep{sh02} have shown that the photon index appears to saturate 
at high X-ray fluxes.  To explain this effect, which \citet{la03} also
observe in NGC~4051, \citet{imhpap98} and
\citet{sh02} independently proposed a two-component spectral
model, consisting of a soft component with constant spectral slope but variable flux
and a hard component with constant flux, so that at high fluxes the
soft component dominates the spectrum and the spectral slope saturates
to the slope of the soft component.  This model naturally explains the
observed spectral variability as a result of the relative changes in
flux of the hard and soft continuum components, without recourse to
any intrinsic change in the shape of either spectral component.
Using time-resolved spectral analysis of a long XMM-Newton observation
of MCG-6-30-15, \citet{fv03} have shown that, on 10 ks
time-scales, the X-ray spectral variability of this AGN is accounted for
by a two-component model where the soft varying component is a power-law
(with roughly constant slope) and the hard constant component is produced
by very strong reflection from a relativistic disk.  \citet{fv03} suggest that the large amplitude of reflection may be boosted by
gravitational light-bending effects close to a Kerr black hole.

As an alternative to the two-component model, \citet{zdz03} have 
suggested that the apparent saturation of the photon index
at high fluxes may instead be a result of a simple pivoting of the X-ray
spectrum about some relatively high energy ($\sim60$~keV or greater), so that
the photon index increases with the logarithm of the flux.  In order
to understand the physical implications of spectral variability in
Seyfert galaxies, it is
important to distinguish between the spectral-pivoting and two-component
models.  Here, we present a  model
independent technique which enables us to
determine conclusively that the two-component model for spectral
variability is correct in the case of \mgc and the Seyfert galaxy
NGC~3516, but not in NGC~4051 which shows evidence for pivoting.
This technique is outlined in
Section~\ref{sec-method}.  The data collection and reduction is
described in Section~\ref{sec-data}.  The observational results and
the implications for the interpretations of spectral variability in
AGN are discussed in Sections~\ref{sec-results} and~\ref{sec-disc}.

\section{Method}
\label{sec-method}

Consider the total fluxes measured in hard and soft energy bands
($F_{h}$ and $F_{s}$ respectively).  We can express each flux as the
sum of constant ($C_{h}$ and $C_{s}$) and variable ($f_{h}$ and
$f_{s}$) components.   
\begin{equation}
\label{eqn:FS}
F_{s}=f_{s}+C_{s}
\end{equation}
\begin{equation}
\label{eqn:FH}
F_{h}=f_{h}+C_{h}
\end{equation}
A simple parameterisation of flux-dependent spectral variability relates the hard and soft
variable fluxes, by $f_{h}=kf_{s}^{\alpha}$, where $k$ and $\alpha$ are
constants.  If $\alpha<1$ this relation describes power-law pivoting
spectral variability, where the varying power-law pivots about some
energy $E_{\rm p}$ (larger than the observed energies)
while keeping the flux density at $E_{\rm p}$
constant (see appendix in \citealt{zdz03}, and
Fig.~\ref{fig:pivotsim}).  As pivot energy increases, so $\alpha$
asymptotically approaches 1.  In the case of $\alpha=1$, the spectral shape of the varying
component is constant with a hardness ratio $k$.  Values of $\alpha>1$
correspond to a pivot energy which lies below the soft energy band.  Incorporating
varying and constant continuum components, we obtain the following
relation between $F_{h}$ and $F_{s}$:
\begin{equation}
\label{eqn:fhfs1}
F_{h}=k(F_{s}-C_{s})^{\alpha}+C_{h}
\end{equation}
By plotting \fluxplot (which we call a `flux-flux plot'), we
reveal the nature of the spectral
variability.  Specifically, if the varying component has constant
spectral shape
the \fluxplot relationship will be linear with gradient
$k$ (i.e. the hardness ratio of the varying component), and an
intercept $C=C_{h}-kC_{s}$ (this linear form occurs irrespective of the
actual spectral shape of the varying component). In contrast, if the spectral shape of the
varying component depends on flux the \fluxplot
relationship will not be linear.  In particular, if the spectral
variability is explained by simple power-law pivoting, with no constant
components, the \fluxplot relationship will be well fitted with a
simple power-law.  Other non-linear shapes of the \fluxplot
relationship might be envisaged for more complicated forms of
spectral variability, or if the varying component spectrum is not a power-law.
Note that such an approach has
previously been used by \cite{chu01} to show the existence of a
constant soft spectral component in the high/soft state spectrum of the black hole
X-ray binary Cyg~X-1.

\begin{figure}
\begin{center}
{\epsfxsize 0.8\hsize
 \leavevmode
 \epsffile{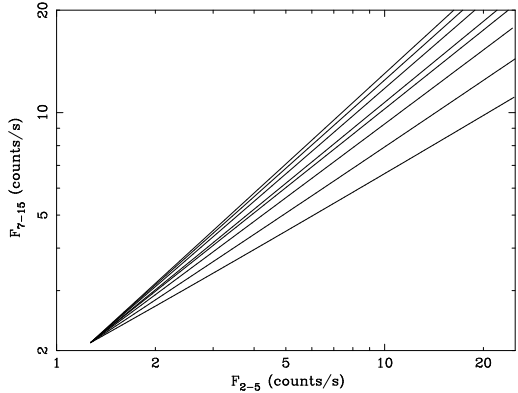}
}\caption{Flux-Flux representation of spectral variability as a
result of spectral pivoting for a number of pivot
energies (from the bottom $E_{p}=30,100,300,1000,3000,10000$~keV with
\fluxplot relations described by power-laws with 
corresponding slopes $\alpha=0.56,0.73,0.79,0.83,0.86,0.88$).  The flux-flux
relations were simulated using the {\sc xspec} {\sc fakeit} command to
create fake \textit{RXTE} spectra, from which the fluxes in each band
can be determined.}
\label{fig:pivotsim}
\end{center}
\end{figure}

\section{Observations and data reduction}
\label{sec-data}

In order to test our method,
we first use PCA data from \textit{RXTE} observations of \mgc to examine the
spectral variability of this source using flux-flux plots.  \mgc is an
ideal test candidate as a two component model has already been suggested to
explain the spectral variability of this source (\citealt{imhpap98};
\citealt{sh02}; \citealt{fv03}).
We combine data from monitoring observations \citep{utt02} and a
$\sim300$~ks long-look observation \citep{lee99}, all
obtained during \textit{RXTE} gain epoch 3.
We extracted 256s-binned light curves for the entire data set in two energy
bands, \mbox{2-5 keV} (soft, channels 0-13) and \mbox{7-15 keV}
(hard, channels 19-41), using Standard 2 data from layer 1 of PCUs
0,1,2 and standard Good Time Interval selection criteria
\citep{la03}.  The latest L7 background models for faint sources were used to determine
the background level for the PCA.

\section{Results}
\label{sec-results}

\subsection{Flux-Flux plots}
\label{sec-fvf}

\begin{table*}
\caption{Results of the linear and power-law fits to the flux-flux
plots for each of the Seyfert galaxies in our sample.  \label{fvffits}}
\begin{center}
\begin{tabular}{lcccccc} \hline
Object & \multicolumn{3}{c}{Linear model} &
\multicolumn{3}{c}{Power-law model} \\
 & $k_{lin}$ & $C$ & $\chi^{2}$/d.o.f. & $k_{pow}$ & $\alpha$ & $\chi^{2}$/d.o.f. \\ \hline \hline
\mgc & $0.516\pm0.006$ & $1.310\pm0.036$ & $11.12/14$ &
 $1.251\pm{0.030}$ & $0.707\pm0.013$ & $34.94/14$ \\
NGC 3516 & $0.598\pm0.008$ & $1.813\pm0.047$ & $13.45/14$ &
 $1.671\pm0.040$ & $0.662\pm0.014$ & $36.98/14$ \\
NGC 5506 & $0.822\pm0.022$ & $1.429\pm0.126$ & $10.62/14$ &
 $1.369\pm0.070$ & $0.850\pm0.020$ & $13.46/14$ \\
NGC 4051 & $0.502\pm0.008$ & $0.452\pm0.022$ & $63.18/11$ &
 $1.040\pm0.003$ & $0.642\pm0.022$ & $22.95/11$ \\ \hline
\end{tabular}
\end{center}
\end{table*}

\begin{figure}
\begin{center}
{\epsfxsize 0.8\hsize
 \leavevmode
 \epsffile{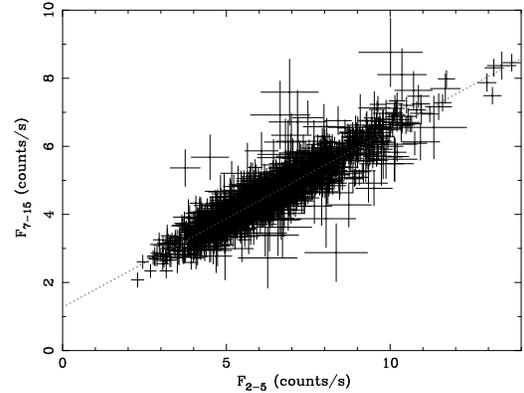}
}\caption{The flux-flux plot for \mgc (see text for details).  The dashed line
represents the linear fit to the data}
\label{fig:fvfunbin}
\end{center}
\end{figure}

Using the 256s-binned hard and soft light curves, we made a flux-flux
plot of the hard flux against the
corresponding soft flux for each 256s segment (Fig.~\ref{fig:fvfunbin}). A linear relationship between the soft and
hard flux is apparent from visual examination of
Figure~\ref{fig:fvfunbin}.  We next fitted a simple linear model,
\linmo, to the data.  We obtained fit parameters
of $k=0.538$ and $C=1.170$ but the fit is poor
(acceptance probability $\ll 0.1$\%), due to significant scatter in the values of $F_{h}$
for a given $F_{s}$.  However, the scatter is small and contains no systematic deviations from
a linear model, implying some weak spectral variability \textit{which is uncorrelated with flux}.

\begin{figure}
\begin{center}
{\epsfxsize 0.8\hsize
 \leavevmode
 \epsffile{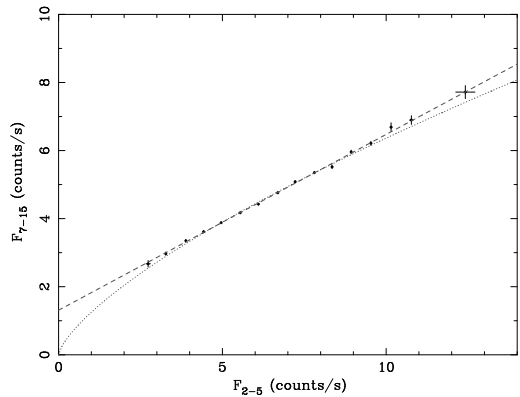}
}\caption{The binned flux-flux plot for \mgc.  The dashed line
represents the linear fit to the data and the dotted line the
power-law fit (without a constant component).}
\label{fig:fvf}
\end{center}
\end{figure}

The nature of the \fluxplot relationship can be seen with greater clarity by
binning up the \fluxplot plot (Fig.~\ref{fig:fvf}). The linear model
then provides a  good fit to the data (\chisq of 11.03 for
16 degrees of freedom) for fit parameters of \mbox{$k =
0.516 \pm 0.006$} and \mbox{$C = 1.312 \pm 0.038$}.  We then fit the
data with a simple power-law to determine whether the spectral
variability may be caused by spectral pivoting of a
single component in the abscence of any other spectral components.  The fit is not formally
acceptable (rejected at $P>99.8\%$ confidence).  Fit parameters of both
linear and power-law model fits are summarised in
Table~\ref{fvffits}.  If a constant spectral component is included in
the power-law fit (i.e. we fit the data with Equation~\ref{eqn:fhfs1})
the fit converges to $\alpha=1$, equivalent to the two-component
linear model.  The $90\%$ confidence lower limit for $\alpha$ using
the power-law plus constant component model is 0.9.  This value of
$\alpha$ corresponds to a pivot energy $E_{\rm p} > 10000$~keV which
seems physically implausible since the high-energy cut-off in \mgc is
at $\sim160$~keV \citep{gua99}.  We therefore reject the pivoting
model for the spectral variability of \mgc.

The linearity of the flux-flux relationship clearly implies a varying
soft component with a spectral
shape which is uncorrelated with flux (and which has a hardness
ratio~$\sim0.52$). A positive intercept on the hard flux axis, suggests the
existence of a constant-flux component, which is harder than the
varying component.  We note that the hard component may be weakly
varying, in order to provide the observed scatter in the unbinned
flux-flux plot or, alternatively, the soft component may show
variations in shape which are uncorrelated with flux.
However, for simplicity we refer to these components as the soft
varying component (SVC) and the hard constant component (HCC), which
we identify with the components of the two-component models suggested
by \citet{imhpap98} and \citet{sh02}, to explain the observed
spectral variability of \mgc.  We will consider the specific spectral
shapes of these components in Section~\ref{sec-soft}.

\subsection{Flux-Flux plots for other Seyferts}

\begin{figure*}
\begin{center}
{\epsfxsize 0.9\hsize
 \leavevmode
 \epsffile{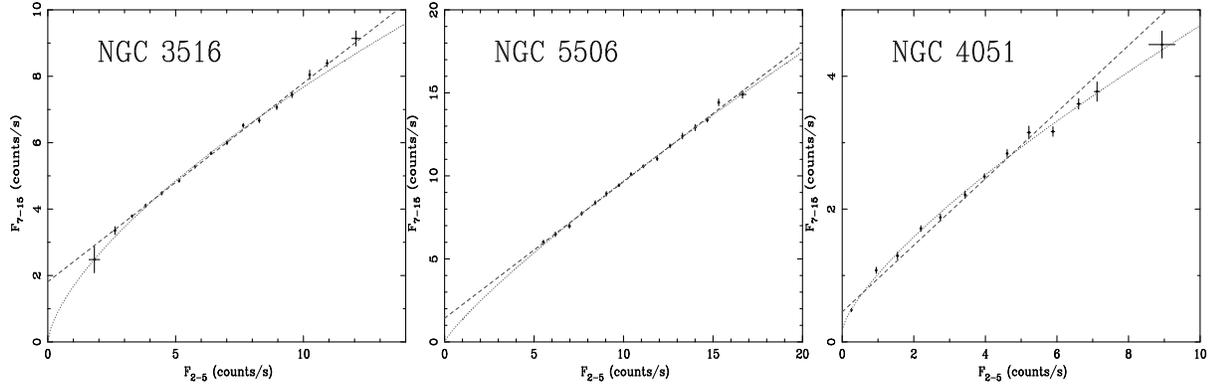}
}\caption{Binned flux-flux plots for NGC~5506, NGC~3516 and NGC~4051.
The dashed lines represent linear fits to the data and the dotted line
the power-law fits (without a constant component except in the case of
NGC~4051 where we show the best fitting power-law plus constant model).}
\label{fig:agnxrbfvf}
\end{center}
\end{figure*}

Having demonstrated that the two-component model for X-ray spectral
variability is correct in the case of \mgc, we now consider the
spectral variability of some other Seyfert galaxies.
We show binned-up flux-flux plots for NGC~5506, NGC~3516 and NGC~4051
in Fig.~\ref{fig:agnxrbfvf}, made using {\it RXTE} monitoring data, in
the same manner as the \mgc flux-flux plot (see \citealt{la00},
\citealt{la03} and \citealt{utt02} for descriptions of the data).  We
fitted all the flux-flux plots with the linear two component model and
the simple power-law model.  The fit parameters are shown in
Table~\ref{fvffits} and the corresponding models plotting in
Fig~\ref{fig:agnxrbfvf}.  The NGC~3516 flux-flux plot is well fitted
by a linear model, but a simple power-law is rejected by the data
(at $P>99.9\%$ confidence).  As in the case of \mgc, the addition of a
constant component to the power-law model yields a value of $\alpha =
1$ (equivalent to the two-component model).   At 90\% confidence the
lower limit for $\alpha$ is 0.86, which corresponds to a pivot energy
$E_{\rm p} > 3000$~keV. Hence we rule out spectral pivoting in the case of NGC~3516.  Like
\mgc the spectral variability in NGC~3516 results from two spectral
components.  The NGC~5506 flux-flux plot can be satisfactorily fitted by both the linear
and power-law models.  This ambiguity probably stems from the narrower flux
range covered by the NGC~5506 data (which is a factor of two lower than
seen in \mgc and NGC~3516).  We note that the best fitting power-law
index to the NGC~5506 flux-flux plot corresponds to a pivot energy
$E_{\rm p}\sim3000$~keV which again we consider unlikely.

A linear fit to the flux-flux plot for NGC~4051 can be
rejected.  The power-law model is a much better fit but is still
rejected at $P>98\%$, although we note that the pivot energy predicted
by this model (around 50-60~keV) coincides with that estimated from the photon-index-flux
correlation by \citealt{zdz02}).  The addition of a constant component to
the power-law results in a formally acceptable fit (\chisq  of 13.38 for 9
degrees of freedom), with fit parameters $\alpha=0.738\pm^{0.035}_{0.055}$
(corresponding to pivot energy $E_{\rm p}\sim300$~keV),
$k_{pow}=0.838\pm0.104$ and $C_{h}=0.183\pm0.01$ for $C_{s}=0$.  A hard
constant component appears to be required but the bulk of the spectral
variability of NGC~4051 is a result of spectral pivoting.

\subsection{The nature of the hard constant component}
\label{sec-soft}

We have shown that there exist at least two modes of spectral variability
in Seyfert galaxies: variability described by a simple linear
two-component model
where the soft component (SVC) varies in amplitude but not spectral shape,
with respect to a hard component (HCC); and spectral variability where the
varying
component pivots, but a hard component is also present.  In either case
the hard component may be constant or weakly varying, but its flux is not
correlated with the flux of the soft component. The SVC may simply
understood as the varying power-law emission component
(e.g. as measured using the `difference
spectrum', \citealt{fv03}). An interesting question is the nature and
physical origin of the HCC.  We now briefly investigate the shape of the
HCC in the case of \mgc, using information contained in the flux-flux plot.

As described in Section~\ref{sec-method}, the intercept on the hard flux axis of the
flux-flux plot, $C=C_{\rm h}-kC_{\rm s}$, where $C_{\rm h}$ and
$C_{\rm s}$ are the fluxes of the HCC in hard and soft bands
respectively, and $k$ is the hardness ratio of the SVC.  In principle,
one could measure many values of $C$ from
flux-flux plots for many energy bands, to create a set of
simultanous equations for the values of $C$ which can be solved to
yield the HCC flux in each band.  However, the resulting equations
(which we do not show here, but are simple to derive) are 
highly sensitive to the measured values of $k$ and $C$ so that 
even small uncertainties in the measured flux-flux relation make this
method unfeasible.  Instead, one can make a simple estimate
of the shape of the HCC by making flux-flux plots in many
bands and assuming the value of $C_{\rm s}$ in the softest band and
then extrapolating the observed linear flux-flux relation to find the
value of hard flux corresponding to that $C_{\rm s}$.

\begin{figure}
\begin{center}
{\epsfxsize 0.8\hsize
 \leavevmode
 \epsffile{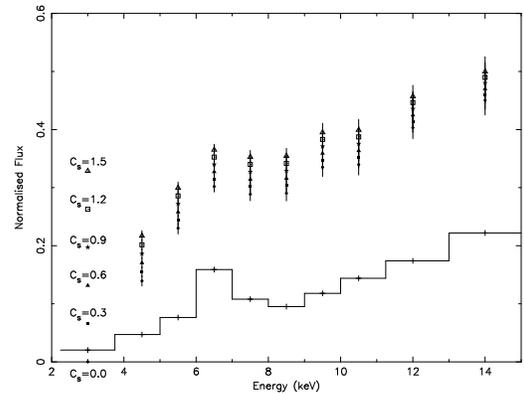}
}\caption{Estimated HCC spectra normalised by the total average
spectrum (see \ref{sec-soft} for details).  Also plotted as a stepped
line is the
expected maximal reflection spectrum (including a narrow iron K$\alpha$ line)
for cold distant material which covers half the sky as seen from the
continuum source (assuming solar abundances and reflector inclination angle of 0$^\circ$).}
\label{fig:hard}
\end{center}
\end{figure}

To implement this method, we measured lightcurves in 10 narrow energy
bands which do not overlap.  The softest band covers the 2-4~keV range, seven bands of width
$\sim1$~keV cover energies from 4-11~keV and two further bands cover
11-13~keV and 13-15~keV.  We then made flux-flux plots for each of the
bands above 4~keV versus the 2-4~keV band.  All the flux-flux plots
were well fitted by linear relations with intercepts on the hard flux
axis.  We chose 6 test values of the 2-4~keV HCC flux, $C_{\rm s}$, in
intervals of 0.3~count~s$^{-1}$ from $C_{\rm s}=0$~count~s$^{-1}$
to $C_{\rm s}=1.5$~count~s$^{-1}$ and determined the corresponding HCC
flux in the other bands using the flux-flux relations (errors are
calculated by propagating through the uncertainties in the linear fit parameters).  In
Fig.~\ref{fig:hard} we show the resulting family of HCC spectra,
normalised relative to the total average spectrum, in order to show
the shape of the HCC independent of the instrumental response.  Note that the shape of the
HCC above 4~keV is only weakly dependent on the assumed values of $C_{\rm s}$, so the
method appears to be quite robust in constraining the HCC shape.  It
can be seen that the HCC is significantly harder than the total
spectrum.  Furthermore, a prominent feature can be seen around 6-7~keV,
suggesting that the HCC contains stronger iron features
than are seen in the total spectrum.

A simple interpretation of the HCC is that it corresponds to
reflection of the SVC from distant cold material (perhaps the putative
molecular torus), and hence can remain constant despite large changes
in the illuminating continuum flux.  However, we note that even the
minimum possible contribution of the HCC (corresponding to $C_{\rm s}=0$) 
exceeds 30\% of the total flux at 10~keV, significantly larger than
the $<10\%$ of total flux expected for reflection from cold material
which covers half the sky as seen from the continuum source
(see Fig.~\ref{fig:hard} and \citealt{gfab91}; \citealt{zm95}). 
We will examine the shape of the
HCC in more detail, and in other AGN,
in a future paper, but for now we concur with the interpretation
of \citet{fv03} that the HCC may represent reflection from the inner disk
which is boosted by light-bending effects close to a rapidly rotating
black hole (although how such a component remains constant is not yet
clear, but see \citealt{fv03} for further discussion).

\section{discussion}
\label{sec-disc}

We have introduced the flux-flux plot as a model independent technique
for distiguishing between two modes of spectral variability in Seyfert
galaxy spectra; pivoting of a single spectral component or changes in
the relative normalisation of a constant hard and varying soft
component with constant spectral shape (the
two-component model).  We have shown that the two-component model is
correct for \mgc and NGC~3516 but the pivoting model applies in
NGC~4051 (albeit with an additional constant hard component).  The
data for NGC~5506 is less clear cut and does not allow a definitive
conclusion as to the nature of the spectral variability in this AGN,
probably due to the smaller flux range sampled by the data.
This ambiguity in the nature of the flux-flux plot can be
avoided by comparing fluxes in more widely seperated hard and
soft bands, so that the index of the power-law flux-flux plot expected
from pivoting is decreased (due to the greater
relative change in the hard and soft fluxes). \textit{XMM-Newton}
observations should be able to provide the necessary separation of the hard and soft energy bands
that is required, to distinguish the two possibilities in NGC~5506 and
similar cases.   

It is interesting that both pivoting and constant-shape modes of
spectral variability are seen in the black hole candidate Cyg~X-1 \citep{zdz02}
with pivoting being observed in the hard state only and a constant
spectral shape seen in both hard and soft states.
\citet{zdz02} review how these different types
of spectral variability can result from a two-phase disk-corona model,
where the accretion disk illuminates the hot corona with low energy seed photons
to produce the hard X-ray emission via inverse Compton scattering 
(e.g. \citealt{haa93}), which in turn feeds back to affect the seed
photon emission by heating the disk.  For example, spectral pivoting results when the
luminosity of the corona (which dissipates most of the gravitational
energy release) is kept constant, while the seed photon luminosity
varies.  Spectral pivoting may also be produced if the corona is pair
dominated.  A constant spectral shape (as in the case of the
two-component model) is produced when the total luminosity of both
disk and corona increases.  \citet{zdz02} suggest that changes in the
seed photon luminosity or total luminosity may correspond,
respectively, to a reduction in disk inner radius (so the corona sees
more of the disk emission) or an increase in accretion rate.  However,
it is not clear how these physical models can explain the very rapid
spectral variations in AGN, where most of the flux variations (and
corresponding spectral changes) occur on short time-scales (hours to
days), much shorter than the expected viscous time-scales in these objects.

Applying the AGN/X-ray binary analogy suggested by the remarkably similar
variability properties of these objects (e.g. \citealt{utt02}), one
might think that the spectral pivoting observed for NGC~4051 implies
that this AGN occupies the hard state, corresponding to a low accretion
rate (few per cent of Eddington).  
However, the X-ray timing properties (power spectrum) of NGC~4051 show
a strong similarity to the soft state of Cyg~X-1 ($\rm M^{c}Hardy$~et al., 2003, in prep), and the low mass of
this AGN suggests a high accretion rate of a few tens of per cent or larger
(\citealt{she03}, $\rm M^{c}Hardy$~et al., 2003, in prep.).  The power
spectrum and spectral-timing properties (coherence, phase lags) of
NGC~4051 $\rm M^{c}Hardy$~et al. (2003, in prep.), are also similar to those of \mgc \citep{vau03}, despite the
apprently different spectral variability modes operating in these
AGN.  

One apparent difference between NGC~4051 and \mgc is that
NGC~4051 shows a significantly larger variability amplitude (by about
a factor 2) than \mgc, on all time-scales.  It is difficult to see how such a large variability
amplitude (especially on short time-scales) could be caused by
variations in the seed photon flux.  However, if the seed photon
luminosity is dominated by internal viscous heating in the disk
(i.e. X-ray heating is not significant) {\it but} the corona
dissipates most of the accretion power, then a
relatively small change in dissipation rate in the corona,
if fed back to the disk, could result in a large change in the seed
photon luminosity and relatively little change in coronal (and total) luminosity,
thus satisfying the conditions for spectral pivoting.  Different
dissipation mechanisms, and different coupling strengths between disk
and corona could then result in the different spectral variability
modes that we see.  The similar power spectral shape and spectral-timing
properties observed in \mgc and NGC~4051 may indicate that these
properties are linked to the more general properties of the corona
(e.g. radial distribution of emission, temperature gradient) rather than the
disk-corona coupling mechanism (e.g. see \citealt{kot01}
for a possible model which can explain these timing properties).

Finally, we note that the hard, constant component in the spectrum of \mgc appears to be
broadly consistent with a pure reflection spectrum but the strength of
this component suggests that the reflection is somehow boosted,
perhaps by relativistic beaming and gravitational light bending
effects \citep{mar00}.  Such a scenario can be realised if the reflector lies close
to a central Kerr black hole,  as also suggested by \citet{fv03} who
describe how such a model might explain the constancy of the
reflection spectrum despite large changes in the soft component flux.

\subsection*{Acknowledgments}
We wish to thank Simon Vaughan for valuable comments and discussions.
This research has made use of data obtained from the High Energy
Astrophysics Science Archive Research Center (HEASARC), provided by
NASA's Goddard Space Flight Center.

\bsp

\label{lastpage}

\end{document}